\documentclass[12pt]{article}

\title{Dialogue on Classical and Quantum between mathematician
and experimenter}

 \author{Andrei Khrennikov\\
 International Center for Mathematical Modeling\\
 in Physics and Cognitive Sciences,\\
 MSI, University of V\"axj\"o, S-35195, Sweden\\ Email:Andrei.Khrennikov@msi.vxu.se\\
Johann Summhammer \\Atominstitut der \"Osterreichischen Universit\"aten \\ Stadionallee 2, 
A-1020 Vienna, Austria\\E-mail: summhammer@ati.ac.at} 

\begin{document}
 \maketitle

\abstract{This dialogue took place at V\"axj\"o, 19 November 2001. 
The main aim of our meeting in V\"axj\"o was to clarify 
our viewpoints on foundations of quantum mechanics.
The most attractive in our discussion was the extreme difference
in our quantum experiences. On one side, pure mathematician
(specializing in foundations of probability theory), Andrei Khrennikov;
on the other side, pure experimenter (specializing in neutron and electron 
interferometry), Johann Summhammer. On one hand, an attempt to test mathematical models
for larger and larger domains of physical reality. On the other hand,
an attempt to create this reality from experimental information -
roughly speaking from clicks of detectors.}

\newpage

Andrei: Was the transition from classical to quantum physics a jump-like
transition that changed the whole picture of physical reality or just a
new application of "classical" statistical physics?

Johann: It was a crucial change of the whole physical ideology.
In any case it was not just a new application of classical statistical
physics.

Andrei: What are the main distinguishing features of quantum physics?

Johann: I would like to underline two new fundamental features:

(D) Discreteness of quantum observations and finiteness of information 
that could be obtained via a concrete measurement. In modern language it 
would be that in finite time only a finite amount of information can be 
obtained.

(R) Randomness of results of measurements.

Andrei: In fact, I do not see so large difference from classical
statistical mechanics. Well, discreteness is the additional
condition. But it is natural from the experimentalist viewpoint:
a finite precision of each concrete measurement. This problem 
was deeply investigated already in 19th century: precision, errors
and so on. Regarding to (R): I do not see any difference with
"classical" statistical physics at all. Already in statistical mechanics
we calculate probabilities and averages. It looks that quantum mechanics
is just a new domain (it might be very special) of applications of
"classical" statistical physics.

Johann: Not at all! First I shall try to explain the role of (D) and then go to (R).
Yes, already
in 19th century (and even earlier) people paid attention to precision of
measurements, errors and so on. However(!), they did not understand the fundamental
role of discreteness of information that could be obtained from measurements.
Suppose you asked an engineer of 19th century: 
"Is it possible to measure the length of this stick with arbitrary precision in finite time?"
The answer would be: "In principle yes. If we use
a very good  measurement  device, we could do this." So for him the problem was
just of a technical matter: we need to improve technology, produce a good device
and perform the precise measurement.

Andrei: I am not sure that an engineer of 19th century would behave in such a way.
First of all as an engineer he should speak about a measurement with an 
arbitrary precision, instead of a precise measurement. Moreover, I think
that he even would not claim that there exists  a kind of "real" length of 
a stick. On the other hand, it seems that you are right regarding to mathematicians
and physicists. In 19th century they were under great influence of 
Newtonian model of reality that was based on the use of "continuous quantities"
(represented by real numbers in a mathematical model). So it might be that
discreteness was surprising for people working with real model of space-time.
From this point of view the (D)-argument is just an argument against the use
of the mathematical model of physical reality based on real numbers. However,
quantum theory did not proceed in this way: the Newtonian "continuous" space-time
was transferred automatically into quantum formalism. So the (D)-argument was not 
so intriguing for experimenters and engineers; it was surprising for
mathematicians and physicists working in the continuous Newtonian model. However, 
in the latter case the (D)-argument did not induce a revolution against
the use of continuous quantities in physics - real numbers. It just modifies the
formalism (by using matrices) to combine discreteness of results of measurements
with "continuous" reality.

Johann: Well, I take this already as an important lesson. 
Observation gives discrete and finite results, 
but the reality which we construe from it is left continuous. This is a hint that the reality 
we can speak about, is the one construed from observations. It is a picture. I think in classical 
physics we were not forced so rigorously to distinguish between observed data and what we 
envision from them. Somehow we could think that the data "are" the world.


Now - randomness. There is a crucial difference between randomness
in classical statistical physics and quantum physics. Classical randomness
is the result of our non sufficient knowledge of objective properties
of physical systems. In classical physics you were permitted to think that each 
classical particle has a definite position at each
instant of time. In principle (at least by improving technology) we could 
find the precise position of a particle. We could not do this for a quantum particle, 
without strongly influencing its future behaviour. This implies that, the image of 
a particle as a small something that moves through space, cannot be upheld.

Andrei: So we could not use the classical probability theory that was used
in "classical" statistical physics? Just to consider a large ensemble
of quantum particles, perform statistical measurement of their properties
and find corresponding probabilities and averages?

Johann: No, we could not do this! Classical randomness is, in principle, reducible
to deterministic objective realistic description. Quantum stochasticity has
a fundamental irreducible character.

Andrei: So classical Ensemble Probability could not work for ensembles
of quantum systems?

Johann: No it would not. You need to work with the complex probability amplitudes of
quantum theory. This has been established in very many experiments. The very working of
semiconductor gadgets, to take just one example, is a consequence of electrons following
quantum probability rules, not classical rules. [Or take everyday chemistry, like paints 
and detergents. If you want to understand why they work as they do, you will not be
able to do this with classical probability applied to electrons and ions. This is sometimes
overlooked when people introduce hidden variable models.

Andrei: What is the origin of such a difference in random behaviour of classical
and quantum particles? Do you think (as many) that there is Something very special
in quantum systems that produces special quantum probability?

Johann: Yes, I think so. 

Andrei: What is this speciality of quantum particles? It may be their wave features?

Johann: 
I am hesitant to use pictures like wave or particle. Although I do it all the time in the laboratory 
and when explaining to students, one must be aware that notions of wave, particle and the like are 
naive projections of our very muscular human experiences onto the world. Although I have recorded 
"waves" in the form of sinusoidal interference patterns and seen "particles" as flashing spots on a 
screen, I think it is wrong to attribute "existence" to waves or particles. Quantum physics forces 
us to recognize that the world simply does not have a "mode of being" such and such. The question 
what something "really" is, is ill-formulated.

Andrei: I also think that the questions of existence and reality are very important
in this discussion. However, I am not sure that these problems are only  related 
to quantum systems. Existence of classical, microscopic, objects is not ABSOLUTE.
It seems that the table in this room "exists" only for a special class of observers
(including human beings). These observers "create" the table by using their
special measurement devices. An observer of the size of an electron or the universe
would never "create" such "an element of reality", table.

Johann: 
From a practical point of view you may be right. But I would argue that any observer, equipped 
with similar reasoning powers as you and I, could find the same invariants and thus the same 
structure, when given the observational data which you and I have, and from which we conclude that 
the particular set of invariants is best subsumed under the notion "massive impenetrable table". 
(Incidentally your data and mine are completely different and yet we believe to be talking about 
the same table.)

Andrei: I think that the crucial role in creation of elements of reality for
a special class of observers plays the relative magnitude of perturbations that
are produced by measurement devices. I still have the idea that there is not such a
large difference between "classical" statistical physics and quantum physics.
These two classes of measurements are characterized by different magnitudes of 
perturbations produced by measurement devices. "Quantum" systems are essentially 
more sensitive to perturbations of our macro-devices than "classical" systems.
Do you agree with such a viewpoint? Negligibly small perturbations - classical
physics, relatively large - quantum. It might be that there exist some classes 
of perturbations that are described neither by quantum nor classical physics.

Johann: 
No, I cannot follow you here. First, I am uneasy with the term 'perturbation'. It creates 
the classical image of one object hitting another one, and this is a very muscular view. 
C.F.von Weizsaecker once said "Hitting and tossing are not 
primitive notions of physics." I agree with him. I would even say that terms like 'scattering' 
or 'interaction' will meet with scrutiny in the future. They are also remnants of classical 
muscular physical thinking. They guide the psychology, which sets the frame for our thinking, 
in the wrong direction.
Aside from that, I think physics can permit only one fundamental description of phenomena. 
Today this is quantum theory. At a coarser level you will, of course, recover the laws of 
classical physics. But this will always be understandable as a washing out of interference 
effects, for instance, when a detector integrates over several interference fringes because 
they are so narrow, or as uncontrollable interaction and thus entanglement with the greater 
environment, so that classical probability rules are sufficient to describe what you observe. 
For practical purposes it may be justified to distinguish between microworld - macroworld. 
But one cannot see this as a limited applicability of quantum theory.

Andrei: Well, it seems that you as well as I identified classical physics (Newtonian
or Maxwellian) physics and classical randomness (ENSEMBLE randomness). I agree
with you that we could not use Newtonian picture of physical reality. I even
do not claim that quantum particles are really particles, i.e., objects localized
in the Euclidean or Minkowski space. These are some objects, "somethings". The only
thing that I defend is that there is no statistical difference between measurements
over ensembles of "classical"  and "quantum" somethings. These are collections of 
somethings that have some internal (well, objective) properties. We perform measurements
of these properties. However, our measurements produce perturbations. This is the 
origin of randomness, both classical and quantum? Of course, quantum and classical
measurements are characterized by different magnitudes of perturbations. But it
is not a consequence of an exotic quantum features. It is just a question of scaling.
An observer who has the size of the universe would produce perturbations of quantum
magnitude by performing a measurement over an ensemble of our
tables.

Johann: No, the different statistical behavior of 'quantum objects', if you permit this term, cannot 
be reproduced by classical behavior, e.g., when a particle is in a superposition of two different
energy states, unless you are willing to introduce an exotic entity like 
Bohm's quantum potential, which violates special relativity left and right.
Moreover, as I just said, I do not like your notion of perturbations.
Newtonian physics could not be used in a quantum framework,
and notions like forces and perturbations seem to me to be a hindrance. Unless one employs 
them in the subtle manner as Feynman did: 
You can think classically, follow each particle through all its conceivable paths, but assign 
each path a probability AMPLITUDE, not a probability. The immediate question -Why the probability 
AMPLITUDE? - has no satisfactory explanation until this day, except for symmetry arguments.

Andrei: Are you against my picture of classical+quantum randomness as 
perturbation-randomness?

Johann: At least your notion of perturbation is not well defined.

Andrei. I agree with you that I have to be more precise with perturbations.
I do not consider perturbations as force-like Newtonian perturbations. 
I am not interested in underlining dynamical model. I speak about 
Statistical perturbations - perturbations of probability distributions.
There are some parameters that determine experimental framework, we can say
{\it context} (the system of physical conditions).
It is natural that by varying these parameters we should
change context. Physical systems (if you like somethings) would react in
a new way to context corresponding to perturbed parameters. We obtain a
perturbation of probability distribution. Such a perturbation could be defined
in mathematical terms, see [1]. Suppose that small variations of contex-parameters
produce negligible perturbations of probabilistic distributions. We call such
experiments - classical. Suppose that small variations of contex-parameters
produce non-negligible perturbations of probabilistic distributions. Such a situation
we have (in particular, see [1]) in quantum experiments.

Johann: My first reaction would be 'No'. On second thought it seems to me that you are aiming at a new 
language of speaking about the quantum mechanical time evolution of a system. Not as a continuous 
transformation of a distribution of probability amplitudes, but as a transformation of 
probabilities. Isn't the Bohmian picture just that?
I am doubtful of the advantage of the distinction 'small perturbation = classical probability' 
and 'large perturbation = quantum probability'. If you think it through for actual situations, 
e.g. the double slit experiment with objects of very small and of very large de Broglie 
wavelength, you may obtain an understanding of what you mean by 'perturbation'. This could be 
interesting. Although I am sceptical that a deeper explanation of 'why the quantum probability 
rule?' can be achieved, I am ready to learn. But how about the superposition of alternatives as 
in the EPR-experiment or the GHZ-experiment? Would perturbation be scaled by how much random spin 
flip there is? Then you would have the opposite of what you want: No perturbation = quantum 
probability, large perturbation = classical probability.

[Andrei: The EPR experiment (as well as GHZ are, in fact, experiments on combination 
of statistical data obtained for distinct complexes of physical conditions, contexts.
EPR-contexts are determined by different settings of Stern-Gerlach devices. 
Here variation of context-parameters, namely parameters of Stern-Gerlach devices,
produces non-negligible perturbations. In fact, Bell's inequality is a kind of constraint 
restricting statistical variations due to  variations of context parameters, see [].

Johann: OK, I see what you mean. The term 'perturbation' tends to lead me astray. It conjures
up a 'hitting-tossing' image. 

Andrei. Recently I read Einstein's papers on blackbody radiation and photoelectric
effect. First of all I was surprised by clearness of presentation of ideas 
(especially, compare to Bohr, Heisenberg, Dirac). So Einstein derived 
"quantum probabilistic distributions" by using methods of classical statistical physics.
These were  the standard statistical manipulations with ensembles of particles +
discreteness. However, the (D)-argument does not look so non classical. OK,
discrete values, but ordinary probability?

Johann: 
In hindsight I would say that Einstein got the right result, because superposition of different 
number states of the electromagnetic field did not play a role in the problem. If it had, as for 
instance for a field in a cavity consisting of a superposition of a 4-photon and a 5-photon 
state and the question is what phase shift will an atom pick up when it passes through, he 
would probably have insisted that there are either really 4 photons or really 5 photons in 
the cavity, if I apply the reasoning he used in the EPR-paper. This would have led to the 
wrong phase shift for the atom.

Andrei: Do you think that Einstein obtained "quantum probabilistic distributions"
just accidentally? Did  he miss the superposition features of quantum systems
just by accident? 

Johann: Yes, it seems so...

Andrei: I think that you would agree that in each 
particular quantum experiment, as a measurement of some fixed physical observable,
we get classical statistical data and we operate with this data in
the same way as with the statistical data obtained in classical experiments.
It seems that the crucial probabilistic difference between classical and quantum
is demonstrated in the interference experiments. Let us consider the two slit
experiment. As I understood, we both do not try to reduce the problem to
wave description. For you: a particle passes just one slit (when both are open)
or in some way both slits?

Johann: 
First, I agree, that any quantum experiment of fixed conditions gives data which follow a well 
known classical probability distribution: binomial, multinomial, Poissonian. The problem of 
understanding arises when you try to make sense of several such experiments in which one or 
several parameters have been changed, and you attempt a classical visualization of 'what the 
particle(s) did'. As to the double slit experiment, it is clear that the particle picks up 
information (=transformation of probability amplitude) along BOTH paths and this determines 
the probability where it will end up on the screen. But again, I am doubtful of this imagery. 
Because, any attempt of verifying where the particle went changes the experimental condition 
to such a degree, that interference vanishes (here you have a case of 'strong 
perturbation = classical probability'). The information, where the particle went, must be 
principally nonexistent in the universe, then you get (maximum) interference. Now, if this 
information is nowhere available, what sense does it make to stubbornly hold on to the 
picture 'but it must have gone somewhere'? I take this lesson to mean that the old and 
taken-for-granted view of the everyday world, which we think to be well defined and always there, 
is wrong and untenable. That, what is there, is of a lot more abstract nature. 
Personally, I tend towards information as the new key concept. But not information about 
'things'. Information about possible values of properties, and sets of properties, which 
are not there 'by themselves', but which WE define on the basis of symmetries. Which 
is almost as good as saying they are there 'by themselves', because the kind of symmetries 
we can think of to categorize observations is not our free invention, but is determined by 
the laws of thought. And these are definitely not our own creation, but are structures 
which we run up against in our minds.

Andrei: I am not against reality as information reality.
We can consider quantum as well as classical objects as just collections of information.
There exist objective laws of evolution and interaction of information;
in particular, the laws of thought.

Johann: I would not use a mechanical idea like 'information interaction', otherwise I agree.

Andrei: Well, we turn back to probability. I consider interference (classical as 
well as quantum) as "interference" of two contexts or two alternatives. The two
slit experiment for one fixed experimental setting, i.e. both slits are open,
could not demonstrate special "quantum features." Interference fringes by themselves are not
exhibitions of special quantum nature. For instance, we can get interference patterns
in the following classical variant of the two slit experiment. 

Charged macroscopic bodies move through charged screen having two slits. We get a complex 
interference like distribution of bodies on the registration screen. What is the root of this 
interference?
The special character of interaction between charged bodies and charged surface of the first 
screen.
Of course, in quantum two slit experiment we do not have such internal (Newtonian) description
of interaction between a particle and the screen; at least in the conventional quantum formalism.
There is no quantum charge... However, we could use, for example, Bohmian mechanics (pilot wave
formalism) with its quantum potential or quantum field given by the wave function.

Johann:
Yes, you can get such periodic-looking classical particle scattering patterns in certain classical 
experiments. But this has nothing to do with quantum interference. As to Bohmian mechanics, I view
this as a perhaps desparate attempt to save our classical intuitive pictures. What I criticize is
that it misses the message of quantum theory, which is similar to the message of special and general
relativity theory. Namely that, our classical intuition, in which the world is made up of distinct 
things, which exercise forces on each other, and all this happens in a well-set three-dimensional 
expanse, is simply wrong. I read the message nature gives us in these theories as: "Thou shallst
not make an image of mine!", a warning, which is probably already in the Old Testament. If we
teach students the Bohmian mechanics, we may help them in the first few steps into quantum theory,
but the damage we do by distracting them from wider schemes of thought is in the
long run much greater. Today there exist computer programs which let you view the temporal 
evolution of the probability density obtained from the Schr\"odinger equation for one or several 
particles for many different Hamiltonians. This can help to form a new kind of 
intuition.

Andrei: In fact, I do not claim the Bohmian mechanics provides the right 
internal description of quantum phenomena. But at least this is one of possible models.

Johann:
Models are certainly needed to do practical research. But I would emphasize seeking 
in data the symmetries, sets of symmetries, sets of sets of symmetries, and so on.

Andrei:  We now turn back to the quantum two slit experiment. So interference by itself is not the 
crucial exhibition of QUANTUM. Troubles with classical picture, in fact, classical probability,
start when we play with a few experimental settings (contexts) in the two slit experiment:
$(C_{12})$ both slits open; $(C_1)$ only the first slit is open; $(C_2)$ only the second slit is 
open. If 
we add probabilities obtained for contexts $(C_1)$ and $(C_2)$ we do not get the probability for
the context $(C_{12})$. 
Classical rule for the addition of probabilities of two alternatives (first or second slit):
\begin{equation}
\label{R1}
P=P_1+P_2.
\end{equation}
is violated. In experiments we get so called quantum rule for addition of probabilities:
\begin{equation}
\label{R2}
P=P_1+P_2+2\sqrt{P_1P_2}\cos \theta.
\end{equation}
Many people consider this fact as the real trouble with classical probability.
They start to think that there is something wrong with the image of electron passing through 
both slits. Moreover, this is the root of the wave approach to quantum mechanics. 

However, we will be in trouble only if we forget about perturbation of probabilistic
distributions induced by the transition from one context to others. In fact, there are
three different contexts that are involved in the consideration: $(C_{12})$ and $(C_1)$, $(C_2)$. 
"Quantum systems" are sensitive to variations of context parameters; in our case:
opening and closing of slits. For "quantum systems" the complex of physical conditions
$(C_{12})$ differs strongly from complexes $(C_1)$ and $(C_2)$. In fact, (\ref{R2})
is just the transformation that connects probability distributions related to different
contexts. The classical rule (\ref{R1}) was derived for the fixed context -
in conventional axiomatics of probability theory, Kolmogorov, 1933. 
Thus if we control context dependence in the right way, then we shall not obtain
any contradiction between quantum statistical data and classical Kolmogorov probability
theory. The main problem in the right probabilistic understanding of quantum 
statistical experiments is that people try to apply theory of probability that
was developed for one fixed context to statistical data obtain in a few different 
contexts.

Johann:
When you speak of contexts rather than perturbation, I can follow you more easily.
Because then you sound like Niels Bohr, who insisted that the whole experimental
arrangement determines the probability distribution for the possible outcomes. He is absolutely
right, if you consider that the slightest change in the experiment must be entered as a 
corresponding change in the Hamiltonian which goes into the Schr\"odinger equation.

As to the Kolmogorovian axioms of probability theory, you seem to imply that quantum theory does
not contradict them. I agree with you. The particle was sent onto a diaphragm with two slits and
later ended up at some point on the screen. This is only one event, observed under one context, not
an event that can have happened under two different contexts. This is not possible. Any given 
experimental situation is always only one context. Nevertheless, the task remains, how to explain the
existence of a lawful relation between the probabilities observed in the three different contexts.
(And jokingly one might ask, why should two contexts suffice to tell us something about a third? 
Perhaps it takes three, to fully predict the outcomes under a fourth? Then we would need a 
three-slit experiment to capture the whole mystery of quantum physics.)

I have a way of viewing this, which is based on the idea that, more observations shall permit 
more accurate predictions. This is a natural assumption for an experimenter. But
it leads to strong constraints on how the probabilities observed under two different contexts
can functionally be related to the probabilities found under a third one. In the two-slit case
it means that, if you measure the probabilities for the particle to arrive at a particular spot
on the screen in two different contexts, you should be able to predict the probability, or at 
least a function of it, under the third context ever more accurately, the more trials you have 
done to measure the first two.
If you follow this through, you get a real and a complex way of describing this, and the latter
is the quantum theoretical rule. But you do not get the classical rule [2]. 
 For this reason I am interested 
in questions like "Which theory permits the most accurate predictions from the information
that is factually available now?" as an entry point to understanding quantum theory. 

Maybe I can comment a little more on the relation between classical and quantum probabilistic
behavior. I think at bottom there is only quantum probability.
Classical probabilistic behavior appears because one does not look closely, and so
collects 'dirty' facts, either by not controlling the context sufficiently, so that the
observed data are actually collected under many different contexts, or by resolving
the measurement results with insufficient fine grading. Let us take the double slit experiment
again. If you do this experiment with an ensemble of particles which do not all have the same
energy, but a wide spread of energies, the probability distribution on the screen will look 
very much as if you had used classical particles of the same wide energy distribution. So you
have actually averaged over many different contexts, because every specific energy requires 
a different preparation context. On the other hand, insufficient fine grading
of the measurement would mean, that you have interference fringes on the screen, which are narrower
than the grains of the emulsion of the photographic plate, or the size of the individual
pixels, if you use a modern digital array chip as detection screen. I have no doubt that all 
phenomena observed in classical probabilistic physics can be traced back to 'sloppy' context 
definition and 'sloppy' data registration, and mostly to a combination of both. And the probability
rule $P_1 + P_2 = P_{12}$ works there only, because the sloppiness of experimentation is 
compensated with a really wrong application of the notion of 'event', as it appears in 
Kolmogorov's third axiom. Of course, in classical physics both mistakes were done unknowingly. I
consider it an accident that the classical rule works in the everyday world, rather than a 
self-evident truth.

Andrei: It seems to me that in fact the discovery of quantum formalism was merely
a discovery in probability theory and not really in physics. Quantum formalism,
at least Hilbert space calculus, is a new mathematical theory to work with statistical
data belonging to a few different contexts. Moreover, it was discovered that nonlinear
probabilistic transformation (\ref{R2}) can be represented as a linear transformation
(for square roots of probabilities) in a complex Hilbert space. Typically the main attention
is paid to this Hilbert space calculus. But I think that the crucial point was 
the derivation (at the beginning purely experimental) of transformation (\ref{R2})
connecting probabilities with respect to three different contexts. In fact, linear algebra
can be easily derived from this transformation. Everybody familiar with the elementary geometry
will see that (\ref{R2}) just the well known $\cos$-theorem. This is the rule to find
the third side in a triangle  if we know
lengths of two other sides and the angle $\theta$ between them:
$$
c^2 =a^2 + b^2 - 2 ab \cos \theta\;.
$$
or if we want to have "+" before $\cos$ we use so called {\it parallelogram law:}
\begin{equation}
\label{P}
c^2 = a^2 + b^2 + 2 ab \cos \theta\;.
\end{equation}
Here $c$ is the diagonal of the parallelogram with sides $a$ and $b$ and the angle $\theta$
between these sides. Of course, the parallelogram law is  just the law of linear
(two dimensional Hilbert space) algebra: for finding the length $c$ of the sum ${\bf c}$
of vectors ${\bf a}$ and ${\bf b}$ having lengths $a$ and $b$ and the angle $\theta$ between them.

Johann: I would disagree with the statement that the quantum formalism was not a discovery 
in physics. As I said before, it implied a big change in our approach to understanding 
the world and we have not yet fully explored all consequences. But I can agree with you
on Hilbert space as a new tool for dealing with statistical data.

  References.

[1] A. Khrennikov, Linear representations of probabilistic transformations induced by context transitions.
{\it J. Phys. A: Math. Gen.}, {\bf 34}, 1-17 (2001).

  [2] J. Summhammer, 
  Maximum predictive power and the superposition principle.
  {\it Int. J. Theor.Phys.}, {\bf 33}, 171 (1994).

\end{document}